# Tunable and saturated structural colors from carbon nanotubes incorporated highly ordered photonic opals

*Ravi Shanker[†‡]\*, Izabela Jurewicz[‡] & Alan B. Dalton[₴]\**

[†]Laboratory of Organic Electronics, Linköping University, SE-601 74 Norrköping, Sweden

[‡]Department of Physics, Faculty of Engineering & Physical Sciences, University of Surrey, Guildford, GU2 7XH, UK

[₴]Department of Physics, University of Sussex, Brighton, BN1 9RH, UK

\*Email: *ravi.shanker@liu.se*, a.b.dalton@sussex.ac.uk

## Abstract

Synthetic opals, based on self-assembly of polymeric nanoparticles generally produces fainted/pale structural colors due to too many lattice flaws in the structures. Here we produces carbon nanotubes (CNTs) incorporated high quality 3D photonic opals (PC-CNT) by evaporative self-assembly. Although the CNTs make up only 0.01% of the fabricated photonic opal, their controlled incorporation has dramatic effect to change the color of the photonic crystals from milky white to intense red. Microscopic study suggest that CNT incorporation did not affect the lattice ordering of the photonic crystals. The tunability of structural colors, as a function of incident angle, were tested and varied against Bragg-Snell law. Furthermore, we tested mechanochromic sensing of the photonic opals, demonstrating their potential as visual indicators. This tunable PC-CNT brings many possibilities including strain sensing or structural health care monitoring, as well as being of fundamental interest.

1. Introduction

Developing materials based inspiration from those found in Nature opens the way to innovations that are often unattainable using conventional approaches. One such exciting topic is structural coloration (SC) arising from interaction of light with submicron-scale periodic structures.[1, 2] Brilliant examples of SC can be found in Nature, from the animal kingdom to plants. Remarkable examples include the striking colors of different butterfly wings, Chameleons, Pollia condensata fruits and various beautiful bird feathers.[3] These systems have attracted great interest in recent years, where researchers use natural design as their inspiration to mimic and fabricate nanophotonic structures.[4]

The coloring mechanism of SC is highly different from that of chemical or pigment-based coloration. SC arise due to light interacting with ordered or quasi-ordered nanostructures (e.g. through absorption, reflection/transmission and scattering), whereas chemical or pigment-based colorations involve photoelectrical energy consumption and conversion.[5, 6] Pigment-based coloration or conventional coloring agents have a variety of disadvantages. Importantly, organic dyes tend to fade over time while inorganic pigments are often based on toxic heavy metals.[7] These all have harmful effects on the environment and human beings. By contrast, SC made of dielectric particles (e.g. silica, polystyrene) is free from photochemical degradation and the colors are purer because of the comparable length scale between their structural spacing and wavelength of light.[8] The analogous to natural opals are synthetic photonic opals fabricated by using polymeric microspheres[9] with different methods such as evaporation assembly, vertical deposition, capillary deposition etc.[10-14] Earlier studies suggest previous efforts producing synthetic opal have serious issues including low quality opal fabrication, poor color saturation, rigidity, and milky white appearance, which limit its potential for variety of applications. To overcome this researchers incorporated absorbing nanomaterials such as carbon black and gold into structure to better absorb the whitish colors produced by broadband light scattering.[15-18] However, these methods produce colors with poor hue. Thus, a method for creating vivid structural colors remains eagerly anticipated. A particular interesting candidate is single walled carbon nanotubes (SWCNT) as a filler in the polymeric structure due to their excellent optical properties.[19, 20] However to achieve a homogenous and controlled distribution of these fillers within a polymer matrix is an obstacle commonly encountered.

Building upon our group's previous research on fabricating ordered colloidal assemblies containing graphene as filler[21], we have developed a cost effective evaporation based self-assembly to produce highly ordered photonic structures containing pristine SWCNT. Stable segregated SWCNT dispersion were prepared using non-ionic surfactant combined with rigorous ultra-sonication and repeated centrifugation. Using latex polymer as template opal and segregated SWCNT as a filler we fabricated highly quality photonic opals (PC-CNT) reflecting vivid structural colors. Furthermore, resulting PC-CNT display different tunable structural colors, when stretched, the lattice spacing between the latex spheres, causing PC-CNT to change color and upon releasing, PC-CNT returns to its original color, mimicking chameleon skin property. Our work can pave a way make PC that are cheaper to produce large on large area, and opening up applications from photonic pains, banknote security, flexible visual sensors, and smart clothing etc.[22-25]

## 2 Results and discussion

## 2.1. Evaporation based self-assembly of carbon nanotube incorporated photonic opals

In the present work used latex polymer synthesized by emulsion polymerization (see experimental section for details) with the solids content is 55 wt %. When latex is subject to evaporation based self-assembly, solvent evaporation brings the particles close and form highly close packed structure.[26] Resulting self-assembled opal structure shows bright structural colors. Figure 1a shows schematic illustration of evaporation-based self-assembly process photonic opal and photograph of the initial wet dispersion of the PC-CNT to fabricate photonic opals. In evaporation-based self-assembly, dispersed latex colloids are permitted to settle inside the container as illustrated Figure: 1a. The higher density of suspended particle is the driving force in contrast to surrounding medium. There are two different competing forces, that are present during sedimentation; thermodynamic (i.e. Brownian motion) and gravitational. For smaller particles (< 500nm), thermodynamics plays an important role, whereas gravity is the dominant force for larger particles.[27]

Inset of Figure 1(a) shows the real dispersion in the process of making crystals via evaporation-based self-assembly. One can also observe three different phases called water, suspension and sediment. Figure 1 (b) shows the fabricated optical images pristine (bottom image - whitish and opaque crystal) and PC-CNT (top image red - upper bottle) photonic crystals at normal incidence and tilted at angle of ~ 40°. Under visual appearance it is apparent from the figure 1b PC-CNT shows bright red structural color which gradually changes to green with changing viewing whereas pristine PC shows faint colors.

Isolated SWNT were used in this study which were prepare with wet chemical methods involving ultrasound and non-ionic surfactant Tritonx-100. Figure S1 (a-d) shows the de-bundling process monitored using optical spectroscopy. We observed that SWCNT can be isolated by diluting the dispersion and overcoming surface energy by using surfactant.[28-32] It was observed that average bundle diameter decreases with decreasing concentration and ~0.01 mg/mL is optimal concentration to have larger number of isolated nanotubes. We typically fabricated photonic opal with this concentration. For larger concentration > 0.5mg/ml impossible to make ordered photonic opals as they tend to phase separate and doesn't form the ordered structure. For very low concentration the solid content is dropped down to very low concentration (around 3%) which will takes few month to fabricate photonic opal. We further study the structure hierarchy of the fabricated photonic opals. Figure 1 (c) shows the topographic atomic force microscopic of the ordered growth of the photonic opals. Latex colloids with CNT self-assembled into hexagonal close-packed (hcp) structures and ordering is not disturbed by incorporation of CNT as can be seen by cross-sectional view of the ordered growth of the crystals from microscopic images (Figure S1 e, f). Inset shows FFT well resolved FCC packing in fabricated opal. The Interparticle spacing from the AFM micrograph was found to be 270±5 nm. 3D photonic opals generally are not free from disorder or defects. Most of them are type of vacancies, interstitial, stacking and dislocations also reported by earlier studies.[33-36] By close approximation one can observe in the AFM images there are some of vacancies in terms of missing latex sphere are present in the system. These defects contribute significantly in altering the optical properties.[37] They are also the main source of generating the scattering by changing the direction of propagating wave in the system which can cause inhomogeneous broadening to the photonic bandgap spectrum. As AFM has limitations in terms of area resolution it can probe therefore we further employ Field enhanced Scanning Electron Microscope (FESEM) to determine opal structure. Figure 1 (d) and S1 (f) also reveals long range well-ordered latex sphere with well-defined planes which can be identified as (111) and (100) from into hexagonal close-packed (hcp) structures.

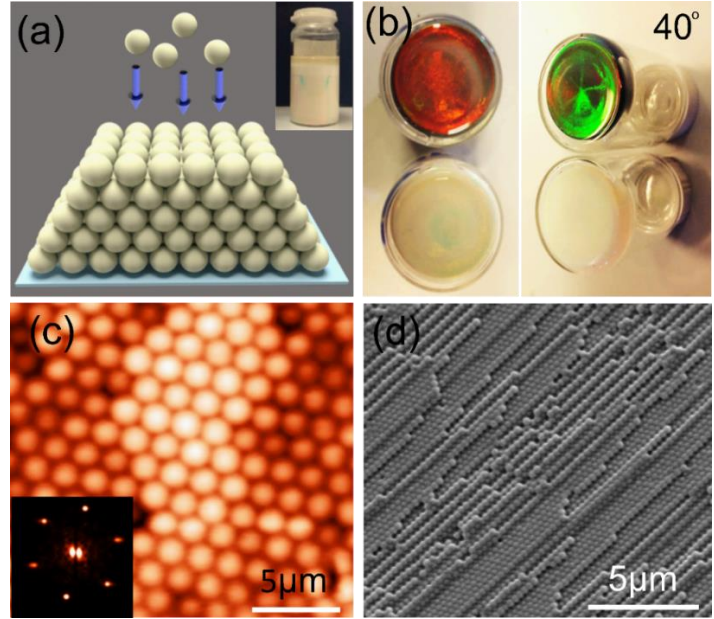

**Figure 1 (a) shows schematic concept of natural gravitational sedimentation process. Inset shows polymer dispersion representing three different phases. Figure 1 (b) represent optical images of photonic crystal with CNT's incorporated (bright reddish - upper bottle) and pristine (whitish and opaque crystal - bottom one) at normal incidence and tilted at angle of ~ 20°. Figure 1(c) represent AFM image of the crystal, inset shows FFT well resolved FCC packing. While 1(d) represent SEM image of the long range ordered growth of the crystals.**

## 2.1. Optical properties of carbon nanotube incorporated photonic opals

Next we study optical properties of pristine opal and CNT infiltrated photonic opal. We investigate the optical properties of the photonic opal by reflection and normal angle transmission and angle resolved transmission spectroscopy method to determine the photonic bandgap. These optical spectroscopic techniques reveal the diffraction or resonance condition at same frequency as a either dip or peak. Photonic bandgap position can be described by the Bragg-Snell law:[38]

$$m\lambda_{max} = 2d_{hkl}\sqrt{(f_1 n_1 + (1-f_1)n_2)^2 sin^2(\theta)} \dots\dots (1)$$

where *m* is the diffraction order, $\lambda_{max}$ is the diffracted wavelength, filling fraction *f* and refractive index *n* of the two media, *θ* is the incidence angle with respect to the normal, $d_{hkl}$ is the interplanar particle separation. The effective refractive index was estimated to 1.34. Based on the effective refractive index and equation (1), we built the diagram of the reflected wavelength from latex polymer size and angle of incident light (Figure 2a), which gives the reflected red color for latex polymer used in this study at normal incidence angle. Figure 2b shows the diffuse reflectance of the PC and PC-CNT samples. The position of the photonic stop band for PC is at approximately 582 nm, whereas the photonic stop band is located at 596 nm in the case of PC-CNT crystal. Importantly, highly ordered PC-CNT crystal led to interesting optical properties reflecting optical signatures of both components. CNT led to selective absorption of light in green-blue part of the spectrum, led to highly saturated and vivid red structural colors compared with pristine crystal. To illustrate that colors form PC-CNT are saturated, reflection spectrum were converted to plotted in the Commission Internationale de l'Eclairage (CIE) chromaticity value as shown in Figure 2c. It is found that PC-CNT produces improved color hue (red square, Figure 2c) and creates high-contrast color compared with PC (orange circle, Figure 2c). Even though these photonic opal look homogenous, they consist of defects, vacancy, and grain boundaries which can affect the optical properties. Therefore, angle-resolved transmission was used to characterize the photonic opals since it probes the whole thickness of the samples and offers rich optical information.[39] Transmission measurement were carried out with 5/nm spectral resolution from Cary 5000 spectrophotometer, the spot size of the light beam was about 2mm$^2$. To rotate the photonic opal about vertical axis an in house develop mechanical rotational stage was used; which allowed an angular step of 5°. Figure 2(d) shows pristine and PC-CNT shows distinct photonic bandgaps with lowered transmission for PC-CNT due to absorptive nature of CNT. The appearances of photonic stop bands are due to Bragg diffraction of the waves.[40] The observed slight red shift in the bandgap position in the case of PC-CNT is due to increased refractive index of the photonic opal than to change in interparticle separation. The noticeable asymmetric transmission spectral shape across the bandgap is due to the different filed distribution on either side of the photonic bandgap and available disorder/imperfection in the photonic opal which induce wavelength dependent Mie scattering in the crystal. The absorption in the short wavelength is higher than the longer wavelength resulting in asymmetrical transmission shape. In reflection spectrum it's not obvious change as it only accounts couple of hundred layers in contrast to whole crystal thickness in transmission measurement, where chances of probing more defects are very strong.

Role of determining the photonic bandgap position is strongly depend upon the propagation of incident wave and axes of the crystals. An incident wave can couple only those Eigen modes of the photonic crystals which possess the same vector k in this situation, changes in the photonic bandgap position with the angle of incidence are equivalent to probing different direction in the FCC Brillouin zone (BZ) or scanning the BZ from L point towards different symmetry points such as W, U, K and edge of the Brillouin zone.[41] The experimental dependence of incident angle with respect to photonic opal are presented in the Figure 2e. With increasing angle of incidence (or moving away from (111) direction) the position of the photonic bandgap shifts its spectral position to the higher energy and gradually become less pronounce. Notably although Bragg peak becomes less pronounce with increasing angle in spectroscopy, the PC-CNT crystal still shows vivid structural colors at higher angle with saturated colors. This is perhaps due to increased index contrast by incorporating broadband absorber into crystal and reducing inhomogeneous background scattering.[21] Next we plotted photonic bandgap positions with varying incident angle and fitted to the Bragg's law (Figure 2f) to obtained effective refractive index and inter-particle separation:

$$\lambda_{hkl}^2 = 4d_{hkl}^2 \left[ n_{eff}^2 - \sin^2\left(\arcsin\left(\frac{\sin\theta_{int}}{n_{eff}}\right)\right) \right]$$

where $d_{hkl}$ and $\theta_{hkl}$ are inter-planer distance and angle between (*hkl*) planes and incident light inside the photonic crystals. To calculate the effective refractive index ($n_{eff}$) is often calculated by volume average refractive index of each material. Fitted equation results in interparticle separation ~ 270 nm and total refractive index 1.34 which suggests the decrease in the interparticle separation ~ 20 nm. This is because of deformation of the latex colloids during solvent evaporation and deformation which is well in agreement with atomic force microscopic and scanning electron microscope results.

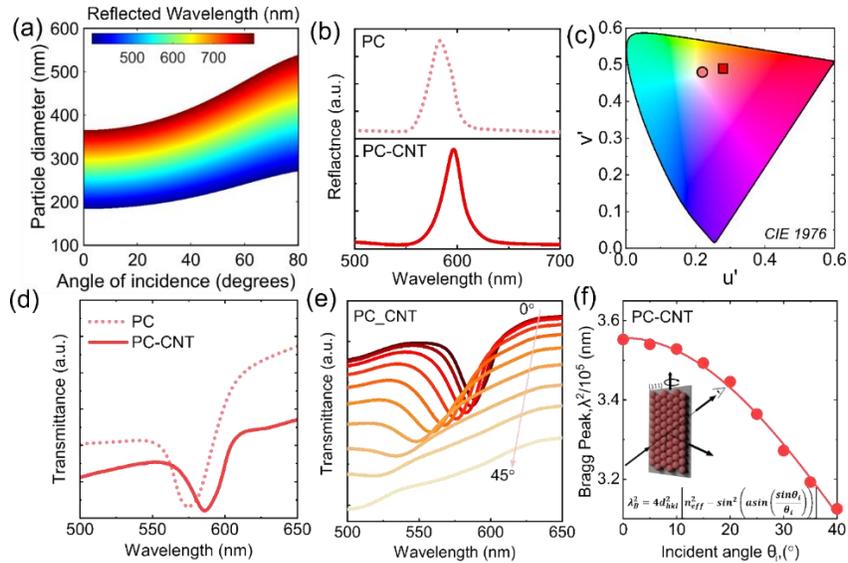

**Figure 2 (a) Graph of reflected wavelength as a function of particle size and varying incident angle. (b) Diffuse reflection spectra of pristine (PC) and CNT filled (PC-CNT) crystals. (c) Representation of reflectance spectra on standard CIE 1976 chromaticity diagram for PC and PC-CNT crystal. (b) Transmittance as a function of wavelength obtained at θ= 0° for a Pristine and PC-CNT, showing significant red-shifting of the photonic band gap position due to the inclusion of CNT. (c) Variation in the transmission spectra with the angle of light incidence for the PC-CNT (d) Angular dispersion of photonic bandgap position**

**fitted to Bragg law for PC-CNT. Thick line represents the theoretical fit. Bragg equation and schematic shown in the inset (where dhkl is the interplanar spacing, neff is the effective refractive index and θ is the angle of incidence).**

## 2.2. Dewetting in PC-CNT photonic opals

Generally colloidal photonic crystals contains substantial amount of the water in the voids which makes them opalescent.[42, 43] The loss of water with time changes the photonic opal appearance and therefore is connected with water loss. The loss or evaporation of water with the time could perhaps due to the changes in the refractive index and inter particle separation of the whole system which leads to change in the spectral position of the photonic bandgap, are studied herein. To test we measured the time evolution of the optical transmission spectra of the photonic crystal to observe these time-resolved spectral changes. These spectra were recorded with the interval of 5 minutes and are presented in the Figure 3 and S2 which shows extinction versus wavelength at different time intervals as interstitial water evaporates. We considered a variation in two main spectral characteristics: variation in the photonic bandgap peak position, and the gradual change of the water peak at 1917 nm which correspond to vibrational band of water molecule. The water peak is associated with the water content and gives quantitative information of local water content and useful for interpreting and correlating with photonic band gap peak. Notably water and Bragg peak doesn't change its shape. Measured spectra are presented as surface contour plot of extinction versus wavelength and time. The color map in figure 3a shows the variations of the change in intensity and Bragg peak position shift with varying time. Inset shows the cut of the map for some selected times. Over period of the time the observed Bragg peak shift is 4 nm thereafter peak position remains nearly constant. Notably, water peak changes rapidly within first 30 minutes then keep changing at slower rate over the next three hours. As evident from figure 3b ~ 50% water content decreases upto ~ 50% of its original value whereas Bragg peak variation is only 1nm.

Normally one would expect large Bragg peak shift with decreasing water content therefore water should be quite low. To estimate Bragg peak shift related to residual water content in opal we used Bragg's law (Section 1) solving for volume fraction of water with approximation of $n_{eff}$ is related to $f_{water}$:

$$\Delta\lambda = 3.26r(n_{water} - 1)\Delta f n_{water}$$

Thus for maximum shift of 1nm within first hour of drying, the maximum water variation is 0.0064 (i.e. 0.64% of the photonic crystal volume). Since this decrease is 50% of the total fwater so total amount of water in photonic crystal is 0.96%. It appears there is two stag drying process here. As it is also evident from figure 3b that there is an initial constant rate of water loss up which is perhaps due to free evaporation of water from interstitial voids between the particles. After about 30 minutes, this period is followed by a falling rate period in which the rate of water loss is seen to slow down. This is likely to related to the transport of water around or through the bead of latex particles or water molecule bound on the latex particle surface.[44, 45] Figure 3 also suggests there is some amount of water still remains in the photonic crystals, i.e. no plateau region has been observed for this spectral feature.

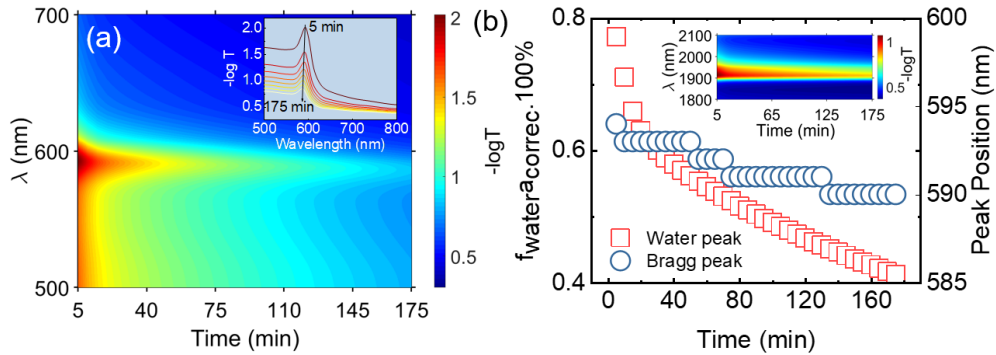

**Figure 3. Time resolved spectroscopy of the photonic bandgap and water peak. 3 (a) Contour plot of the extinction over time and wavelength for PC-CNT. Inset shows cut of the map of temporal changes in photonic bandgap peak for some selected times. (b) The intensity of the water peak has been rescaled using**

**relation $a_{corr}\, f_{water} = 0.96 \times 10^{-4}\,nm^{-1}$, resulting from the estimation of the maximum possible amount of the water in photonic opal. Inset shows time dependent extinction evolution of water peak at 1917nm. Evolution of photonic band gap position (circle) and water absorption peak at 1917nm (square) as function of time.**

## 2.3. Mechanochromic sensing

The ability to tune the optical properties makes PC-CNT attractive candidates for colorimetric sensing. This tuning of the photonic bandgap can be achieved by either changing lattice parameter or effective refractive index.[46-48] Herein we present tuning the stop band by changing the lattice parameter by measuring transmission spectra vs. strain. Sample was fixed on Linkam mechanical stage and in-situ optical measurements have been performed. Figure 4(a) represent changes in the photonic band gap as we increased the strain from 0 to 30%. PC-CNT was readily stretched up to approximately 50% and above. The applied strain impart compression in the direction normal to the plane of the applied strain, resulting in decrease of the photonic opal thickness. The observed blue shift in the photonic bandgap position with applied strain is due to the change in the spacing of the (111) plane separation as shown in in the inset. The shift in the Bragg peak position were plotted as a function of strain resulting in Figure 4b shows that the peak shift is linearly decreased with the applied strain in the visible region. Although the peak shift primarily due to the change in average periodic distance under strain, but it is also affected by several other factors such film thickness change and scattering variations. The fabricated photonic opals are mechanically robust optical sensor and corresponding photographs in relaxed and strain mode in Figure 3c clearly shows that full visible region can be utilized for optical sensing. We further investigate sensitivity of the PC-CNT to the pristine opal. Figure 4b shows fractional change in transmission per 10% increase in strain. PC-CNT shows almost 15 times the sensitivity of the pristine opal, reaching nearly 15% transmission reduction.

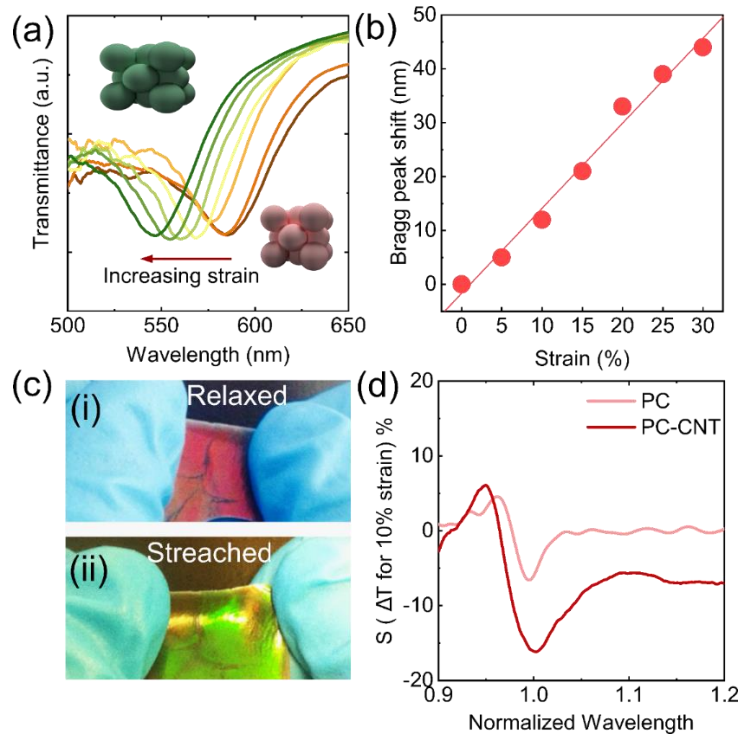

**Figure 4 Sensing application of PC-CNT crystal. (a) Spectral shifting in photonic stop bandgap as function of strain. (b) Plot of the wavelength at the maximum peak shift as a function of strain. (c) Photographs of the PC-CNT optical sensor as a function of applied strain. (d) Sensitivity of the PC-CNT to the pristine opal using fractional change in transmission.**

## 3. Conclusion

In summary, we have demonstrate mechanochromic photonic opal with strong structural coloration. The incorporation of CNT at very low concentration into the interstitial spacing radically enhances the structural colors offering improved visual contrast due to emerging refractive index contrast. These photonic opal offers attractive functional features such as strong iridescence, stretch-tunability, excellent durability and mechanical robustness. Hence these photonic opals are potential candidates for visual calorimetric sensor, photonic paints, photonic display etc.

# Supporting Information

Materials and methods, including preparation of CNT dispersion and characterization, preparation of photonic crystal fabrication, structural and optical characterization. Supplementary figures show de-bundling of CNT by dilution, cross-sectional SEM and AFM images of the photonic crystal, time dependent optical transmission of the photonic crystal.

**Corresponding Authors**

Email: *ravi.shanker@liu.se;*
Email: *a.b.dalton@sussex.ac.uk*

**Author Contributions**

RS and AD conceived the original idea. RS carried out the CNT dispersion preparation, fabrication of photonic crystal, optical characterization, interpreted of the data and wrote the original draft of the manuscript. IJ assisted in preparing CNT dispersion and assisted in SE and AFM imaging. All work was coordinated and overseen by AD. RS wrote the manuscript through contributions of all authors. All authors has approved the final version of the manuscript.

**Notes**

The authors declare no competing financial interest.

**Acknowledgements**

We acknowledge financial support from URS/ORS scholarship from University of Surrey.